# Direct observation of the basic mechanisms of Pd island nucleation on Au(111)

C.S. Casari[*], S. Foglio, F. Siviero, A. Li Bassi, M. Passoni and C.E. Bottani

*Dipartimento di Energia* and

*NEMAS - Center for NanoEngineered Materials and Surfaces,*

*Politecnico di Milano, Via Ponzio 34/3, I-20133 Milano, Italy*

*Corresponding author: Carlo S. Casari mail: carlo.casari@polimi.it

**Abstract**

The formation mechanisms of evaporated Pd islands on the reconstructed Au(111) $22 \times \sqrt{3}$ herringbone surface have been here studied by Scanning Tunneling Microscopy (STM) at room temperature. Atomically resolved STM images at the very early stages of growth provide a direct observation of the mechanisms involved in preferential Pd islands nucleation at the elbows of the herringbone structure. At low Pd coverage the Au(111) herringbone structure remains substantially unperturbed and isolated Pd atoms settled in hollow sites between Au atoms are found nearby the elbows and the distortions of the reconstructed surface. In the same regions, at extremely low coverage (0.003 ML), substituted Pd atoms in lattice sites of the Au(111) surface are also observed, revealing the occurrence of a place exchange mechanism. Substitution seems to play a fundamental role in the nucleation process, forming aggregation centres for incoming atoms and thus leading to the ordered growth of Pd islands on Au(111). Atomically resolved STM images of Pd islands reveal a close-packed arrangement with lattice parameter close to the interatomic distance between gold atoms in the fcc regions of the Au(111) surface. Distortion of the herringbone structure for Pd coverages higher than 0.25 ML indicates strong interaction between the growing islands and the topmost Au(111) layer.

I. INTRODUCTION

The growth of ordered arrays of metal clusters supported on surfaces is of extreme interest both from a fundamental and a technological point of view. Indeed, progress in fields such as model catalysis and magnetic nanostructures is critically dependent on the ability to prepare uniformly-dispersed clusters with controlled characteristics. This permits a precise study of the cluster properties which depend on size, crystalline structure and interaction with the surface, and opens the way to novel applications. The understanding, as much complete as possible, of the basic mechanisms underlying cluster nucleation at surfaces is a key factor in order to achieve these goals.

In this framework the Au(111) $22 \times \sqrt{3}$ surface, displaying the so-called 'herringbone' reconstruction, represents a unique substrate. The periodic arrangement of ridges and elbows, originated by changes in surface domain orientation, sets up a long-range order in the reconstructed surface, which acts like an array of possible preferential nucleation sites for deposited atoms. Island growth nearby the elbows of the Au(111) reconstruction has been observed in the deposition of metals such as Pd, Ni, Co, Pt, Fe, Mo and Ti [1-7]. This preferential growth presents common features such as the position of nucleation sites and island faceting along the close-packed $\langle 1\bar{1}0 \rangle$ directions, but also some relevant differences. First of all, various degrees of intermixing between the surface and the adlayer have been reported, ranging from the case of nickel [8] and platinum [4], for which the growth of islands with mixed composition has been observed at room temperature, to the one of iron [5] and palladium [1], showing nucleation of gold-free islands. Moreover, a variety of different situations are found concerning the interplay between islands and surface during the growth process: such interactions can lead to the onset of island preferential growth along specific directions [9] and to significant distortion or even lifting of the herringbone reconstruction [1], [10]. These and other aspects that will be discussed more in detail in the following clearly indicate that the above mentioned elements experience a diverse interaction with the gold surface, hence differences in the basic island nucleation mechanisms on Au(111) can be expected.

Two different basic processes have been proposed in order to explain island nucleation at the elbows of the herringbone reconstruction. The first one deals with the appearance of lattice sites where gold atoms are replaced by adatoms (i.e. place exchange) to release superficial stress [11]. The sticking probability at these sites should by much higher than anywhere else, rapidly leading to island nucleation. The second mechanism is related to limited adatom diffusion at specific sites nearby the elbows thus favouring island nucleation [1]. Notwithstanding the large number of studies reporting preferential nucleation, for some of the cited elements the occurrence and in particular the actual role of the two mechanisms in the early nucleation process is still not completely clear. Meyer *et al.* [11] proposed the nucleation mechanism based on place exchange suggesting that all elements characterized by higher values of surface free energy and heat of sublimation with respect to Au should show preferential growth.

Controlled deposition of Pd is of great interest in the field of model catalysis because of its remarkable activity in many oxidation reactions [12, 13]. In this framework the formation of Pd-Au surface alloys and the already mentioned ordered growth of Pd on Au(111) have been already reported in a few works [1,14]. According to Meyer et al. [11] Pd belongs to the group of elements for which the place exchange mechanism is expected, even though the values of the surface free energy and heat of sublimation are very close to the gold ones. Anyway, in the case of Pd island growth on Au(111) no direct observation of the nucleation processes has been reported so far. Extensive alloying of Pd islands with gold atoms was proposed as well [15], but its occurrence has been ruled out by STM measurements reported by Stephenson *et al.* [1] who supposed that hindered diffusion across point dislocations may be the mechanism accounting for preferential nucleation at the elbows of the herringbone reconstruction, anyway leaving substantially open the question whether Pd substitution into the Au surface plays a role in the growth process or not.

Here we report on a scanning tunneling microscopy (STM) study of Pd islands evaporated on Au(111). Atomically resolved STM images allow the direct observation of both place exchange between Pd and Au and adatom pinning in fcc/hcp hollow sites nearby the elbows. On the basis of our experimental data we propose the substitution mechanism as the main one responsible for Pd island nucleation at the elbows of the herringbone reconstruction. Finally, we further investigate the island-surface interaction reporting on the observation in successive STM images of mass transport phenomena between neighbouring palladium islands, accompanied by reorganization of the underlying gold surface reconstruction.

## II.    EXPERIMENT

### A.  The Au(111) surface

Before going into the analysis of our experimental data, we briefly recall the main characteristics of the Au(111) reconstructed surface which has been already extensively studied [16-20]. It is originated from an anisotropic compression along the $\langle 1\overline{1}0\rangle$ direction that allows to accommodate 23 atoms per cell in 22 bulk lattice sites, resulting in a $(22 \times \sqrt{3})$ reconstruction (Fig. 1-a). At the same time some Au atoms are pushed aside in $\langle 11\overline{2}\rangle$ direction in hcp sites instead of the usual fcc ones, so that atoms in between the two different stacking positions are forced in higher energy bridge sites and lift up with respect to the surface plane. As a consequence, couples of ridges form that run along the $\langle 11\overline{2}\rangle$ direction, separating hcp regions from fcc ones. Ridges are called *discommensuration lines*, since they represent the transition region from one type of stacking to the other. In correspondence of these boundary lines a periodic domain rotation of ±120° is introduced forming a typical zig–zag pattern often referred to as the ''herringbone'' reconstruction. Two kind of ridges are individuated according to the way they undergo such rotation: the x ridge bulges out towards the convex side of the bend while the y ridge follows a smooth course. As clearly visible in Fig. 1-b, such a difference results in the existence of two different types of elbows named bulged if the prominence of the x ridge is directed towards the fcc region, otherwise named pinched. Discommensuration lines, elbows, hcp and fcc regions all exhibit different degrees of lattice compression with respect to the unreconstructed surface, hence the presence of many non equivalent sites from the energetic point of view must be taken into account.

Actually on a freshly prepared Au(111) - ($22 \times \sqrt{3}$) surface several kinds of defects of the reconstruction are found, as reported comprehensively by Barth *et al* [16]. Such defects are mainly represented by U-shaped connections between adjacent ridges (most containing hcp domains, occasionally fcc ones) or distorted elbows, giving rise to rather irregular patterns sometimes found where zig-zag arrangements with different orientation meet or in the vicinity of terrace steps. Occasionally also holes up to 2 nm in diameter are present on the surface causing irregularities in the reconstruction periodic structure. We also observed regions with pairs of ridges running straight without any elbows for some tens of nm, which we attributed to the interplay between the reconstruction and steps perturbing the usual pathways of surface stress relaxation. In the following analysis of the early stages of Pd island nucleation we will not limit our discussion to a perfect reconstruction pattern, since we noticed that useful information can be gained considering also defected regions, resulting in a more complete picture of the growth process. Moreover, another motive of interest is given by the fact that Pd deposition itself modifies the reconstruction and causes the appearance of irregularities in the surface.

### B. Experimental procedures

Experiments were carried out in a ultra high vacuum (UHV) chamber (base pressure $\approx 5 \times 10^{-11}$ mbar) equipped with standard facilities for sample preparation and an Omicron UHV VT-SPM. The Au(111) surface (a commercial substrate of evaporated gold on mica) was prepared in UHV by 15 min Ar+ sputtering at 1 keV at a sample temperature of 800 K, which was maintained for at least 20 min before cooling to room temperature. A preliminary STM characterization of the clean as-prepared surface was performed in order to exclude contaminations and verify the presence of the ($22 \times \sqrt{3}$) reconstruction.

Pd was deposited at room temperature on the clean Au(111) surface by means of an electron beam evaporation source (Pd wire purity > 99.99%). The deposition rate (~0.035 ML/min) was controlled by monitoring the ion flux, which is proportional to the flux of evaporated atoms. The coverage ranges from 0.003 ML up to 0.50 ML and it was estimated by a software analysis of the STM images. STM measurements were acquired at room temperature after deposition at a sample-tip bias voltage between ± 2 V and a tunneling current in the range 0.1-2 nA using home made etched W tips.

## III. RESULTS AND DISCUSSION

### A. Pd Deposition On Au(111)

The deposition of palladium on the Au(111) surface results in the well-known preferential growth at the elbows of the reconstruction, as shown in Fig. 2 for different coverages from 0.012 ML to 0.14 ML. Islands nucleate mainly just outside the x ridge swelling, i.e. on the fcc region outside bulged elbows and in the hcp region between the ridges of pinched elbows. Occasionally at pinched elbows islands are found also on the fcc region inside the bend, a few sites away from the edge dislocation of the x ridge. Similar observations are made in the case of nucleation on defects (e.g. U connections) of the reconstruction pattern. In analogy with the bulge of the x ridge, U connections containing hcp domains are crossed by an extra row of atoms ending with an edge dislocation at the inner side of the U shape [16], and islands grow on the fcc region right outside there.

Fig. 3 presents a STM image at atomic resolution of an island grown on an fcc domain outside a bulged elbow with the characteristic polygonal shape following the threefold $\langle 1\bar{1}0\rangle$ directions. In this coverage range, islands are one layer high and no features imputable to the presence of gold atoms (i.e. island alloying) are observed. Going deeper into the analysis of island growth dynamics, Fig. 2 shows how the potential nucleation sites, both regular elbows and defects, act like traps which are progressively filled until the island density $n$ (number of islands per unit area) reaches the maximum value $n_{max}$ which represents the density of nucleation traps. This behaviour indicates also that the incident flux is low enough in order not to compromise the ordered growth regime, as shown in different studies regarding dynamics of growth controlled by the presence of traps both by atomistic Monte Carlo simulations [21, 22] and by STM of clusters on Au(111) vicinal surfaces [22] and quasicrystals [23]. Fig. 4 reports island size distributions for coverage ranging from 0.012 to 0.14 ML. As previously found both theoretically and experimentally for different physical systems [22] the curves are well fitted with bell-shaped functions, with mean size and distribution width increasing as a function of coverage. In the pre-saturation regime ($n < n_{max}$), a quite narrow island size distribution is found, while a broader and asymmetrical shape is evidenced once $n_{max}$ has been reached. It is known that symmetrical distributions are observed only when two growth conditions are verified: the first one deals with the diffusion coefficient, which should be high enough to avoid nucleation outside traps. The second one regards the binding energy of atoms in trap sites and the cohesion energy of neighbouring palladium atoms that should be sufficiently high to prevent them from leaving the preferential nucleation sites and to preclude statistically relevant island re-evaporation. Indeed, whenever such conditions are not satisfied, bimodal size distributions have been reported [22].

At low coverage (Fig. 2a-b) we observe that the early stages of growth are not characterized by a uniform occupation of all available traps, since Pd islands with size of some tens of atoms are found in regions where several empty elbows are still present. This can be explained assuming that, once an island has nucleated, it grows rapidly due to its atomic sticking probability, higher than that of an empty trap, giving rise to the observed inhomogeneous island distribution at low coverage. Similar results were reported for the case of both Pd and Ni deposition on Au(111) [1], [2]. In particular in the case of Ni, Chambliss *et al.* [2] have demonstrated that such an island growth scheme is incompatible with a Poisson process in which atoms diffusing on the surface have the same probability of joining an island or pinning at an empty trap, and they have calculated that the sticking coefficient of islands should be at least two orders of magnitude larger. In order to obtain information about the competition between the further growth of islands already present on the surface and nucleation on new sites, in Fig. 5 we report the log-log plot of island density $n$ and mean island size $<s>$ versus coverage $\theta$. In the pre-saturation regime they follow a power-law dependence:

$$n = \theta^{0.72 \pm 0.04}$$
$$\langle s \rangle = \theta^{0.30 \pm 0.05}$$

The sum of the exponents approaches unity, since the product between $n$ and $<s>$ is the coverage, which has been verified to be a linear function of deposition time. This means that in the early stage of growth the major part (about 70%) of atoms deposited on the surface in a given time interval $t_1 - t_0$ is involved in the growth of an island that at time $t_0$ did not exist, while the remaining 30% are involved in the growth of already existing islands. As expected, once the saturation regime has been reached *($n=n_{max}$)*, the slope is nearly one for $<s>$ and zero for $n$.

The layout of islands on the surface, and in particular the occupation probability of different types of nucleation sites, are important sources of information in order to study the nucleation mechanisms, which are the subject of the next section. At low coverage, island occupation of preferential sites does not appear to be random. Actually, the analysis of several STM images reveals that groups of islands form on neighbouring sites or along rows of pinched/bulged elbows while wide regions of the surface including several preferential sites remain uncovered, as visible for example in Fig. 2-a. Quantitative analysis of island disposition on the surface was performed by means of the nearest-neighbour pair correlation index *NNI* [2, 24]. The *NNI* is defined as the ratio of the mean of the nearest neighbour island distances (<*NND$_i$*>) to the mean of the nearest neighbour distances for a uniform distribution of islands (<*RD*>), that is:

$$NNI = \frac{\langle NND_i \rangle}{\langle RD \rangle}$$

The mean island nearest neighbour distance is defined as:

$$\langle NND_i \rangle = \frac{1}{N} \sum_{i=1}^{N} NND_i$$

where *N* is the number of islands. The mean random distance <*RD*> is defined as:

$$\langle RD \rangle = \frac{1}{2} \sqrt{\frac{S_{tot}}{N}}$$

where $S_{tot}$ is the total surface of the investigated area. *NNI* thus gives an estimate of the island disposition on the surface in terms of randomness (*NNI*≈1), repulsive interaction (*NNI*>1) or clustering degree (*NNI*<1) [24]. From the analysis carried out on several STM images for coverage of 0.012 ML (Fig.2a), we found a mean value of *NNI*=0.75, indicating a tendency of the islands to nucleate in neighbouring elbows.

The process of gradual occupation of traps proceeds with different probability depending on the nature of the nucleation sites. Referring to regular elbows, at coverage up to 0.035 ML (Fig.2c), if we consider the same number of bulged and pinched elbows, we observe that the fraction of bulged elbows at which an island has nucleated is slightly higher (about 10%) than that of pinched elbows. More remarkable differences are found when comparing the probability of nucleation on regular elbows and on defects of the reconstruction. At very low coverage islands appear first only on the standard reconstruction pattern (Fig. 2-a), then nucleation proceeds on both kind of sites but with an evident preference for regular elbows. The STM image shown in Fig. 2-b (0.026 ML coverage) is a clear example of this tendency: a consistent fraction of regular elbows is decorated, while regions where zig-zag patterns with diverse orientation meet, producing irregularities in the reconstruction, are almost free from Pd islands. We have calculated that at 0.035 ML an island has grown on about 80% of regular elbows, while less than 50% of U connections and distorted elbows have been decorated. As shown in Fig.2-d, at saturation of island density all elbows or defects have nucleated an island. Such a difference in the nucleation probability suggests that a strong link exists between the pinning basic mechanisms and the local structural order of the surface.

### B. Island Nucleation Mechanisms

We have described the growth of Pd on Au(111) as controlled by the presence of traps (i.e. nucleation sites) on the reconstructed surface, anyway nothing has been specified about their nature yet, i.e. about the basic processes involved in the very early stages of island nucleation. These mechanisms should account for island nucleation in correspondence of the elbows of the reconstruction and for the observed island growth dynamics.

Fig. 6 shows two STM images, collected in succession after deposition of 0.026 ML Pd, showing some bulged elbows where no islands have grown yet. Several depressions (dark spots) are found nearby the elbows, mainly localized on the fcc regions between each pair of ridges, and causing small wrinkling of the surface. These depressions are not observed on the same surface prior to Pd evaporation. We can exclude possible contributions from impurities (e.g. CO) of the evaporant since dark spots do not increase with increasing Pd coverage and completely disappear when Pd islands start to grow. Hence we attribute these features to the presence of Pd atoms adsorbed on the Au(111) surface. A similar chemical contrast effect has been reported for adatoms on Pd(111) [25] and is in agreement with Maroun et al. [14] observing Pd substituted atoms in a Pd-Au alloy surface. It has been also observed in different systems such as Au on Ni(110) [26], V on Pd(111) [27] and In on Si(111) [28]. Pd adsorbed atoms are here always imaged as holes also for different sample-tip bias voltages. During bad tunneling and imaging conditions (i.e. not stable and reproducible images, not reported here) when tip behaviour spontaneously changes, we observed the appearance of slight protrusions (bright spots) instead of holes, probably due to Pd atoms accidentally collected by the tip during the scanning. This clearly indicates that dark spots are not really surface vacancies, but rather Pd adsorbed atoms. At a closer inspection, dark spots are shallow depressions (~0.5 Å) positioned in surface hollow sites as shown in the atomically resolved image of Fig. 7. A sort of cloud surrounding the Pd atom is also visible; such effect may be related with a slight perturbation of the local density of states or with a diffusion of surface states but it is very difficult to establish whether this variation is caused by Au/Pd interaction or by a tip-induced effect. A comparison between the two images reported in Fig.6 reveals that Pd atoms, though already occupying local minimum energy positions in fcc hollow sites, still have a residual mobility around preferential sites. Some of the misfit or missing atoms in the right image have been indicated by an arrow: evident signs of tip manipulation are not present, anyway a role of the scanning process in overcoming activation energy barriers for Pd atom displacement cannot be excluded *a priori*. Here it is important to stress that Pd atom decoration was found for all potential nucleation sites regardless of their nature (regular elbows, U connections etc.), proportionally to the amount of deposited Pd. As useful examples, the image reported in Fig. 8 shows the presence of Pd atoms distributed nearby regular (both pinched and bulged) and distorted elbows. On the contrary, the straight ridges are free from the presence of Pd atoms.

As already mentioned, it has been proposed that the diffusion coefficient of Pd atoms may be very low at edge dislocations at the bulge of the *x* ridge, leading to preferential nucleation at the elbows of the reconstruction [1]. What is actually observed in our STM images is that adatom diffusion is probably hindered in a 20-25 nm wide region surrounding the potential nucleation sites, where Pd atoms collect since the very early deposition stages with a preference for fcc regions, anyway without clustering in close-packed islands. In particular no special tendency to the aggregation on edge dislocations is evidenced. Theoretical studies about the atomic diffusion mechanisms for the Co/Au(111) system [10] and for fcc (111) surfaces in general [29], suggest that the diffusion coefficient on regions of the surface displaying high lattice compression should be lower than on the close-packed fcc-hcp domains. According to Ref. [17] compression is maximum at edge dislocations nearby the elbows, exhibiting three times the value found for discommensuration lines, which in turn display a misfit of about 5.5% with respect to the unreconstructed surface. Actually, Pd atoms do not show any preferential tendency to collect on the ridges, on the contrary a significant fraction of them is found on fcc sites. Therefore, such theoretical considerations about adatom diffusion do not find complete correspondence in what observed experimentally.

The picture emerging from the analysis of our STM data is that hindered diffusion alone does not account for the preferential growth of Pd on Au(111). Since a very similar Pd decoration is common to all nucleation sites, the reaching of a critical size for island nucleation should be equally probable on regular elbows and defects of the reconstruction, in contrast with experimental evidences. Moreover the first stages of growth should lead to a random island spatial distribution, with nucleation on nearest-neighbouring sites being unfavourable, both for statistical motivations and because the growing islands act like very efficient traps for Pd atoms, inhibiting further nucleation nearby. Our findings reveal that Pd atoms collect at the elbows without aggregating and suggest that another mechanism should be involved in the process of island nucleation and growth.

The replacement of Au with Pd adsorbed atoms (place exchange mechanism in the topmost layer) is a thermally activated process employed for the formation of surface alloys as reported by Maroun et. al [14] and in some cases it is already active at room temperature, as previously reported for Pt and Ni [4, 8]. Our STM measurements confirm that such a process is actually occurring at room temperature also in the case of palladium. At a very low Pd coverage (0.003 ML), the atomically resolved image reported in Fig. 9 shows a rather different situation with respect to Pd atoms adsorbed in hollow sites of the close-packed gold surface shown in Fig. 7. Here a depression (0.1-0.2 Å) is found including three neighbouring atomic sites (see Fig.9), which are encircled by gold atoms protruding 0.1-0.3 Å from the surface mean level. The depression is compatible with the presence of atoms in registry with the fcc lattice, as shown by the line profile reported in the figure.

We interpret the STM data as an evidence of the substitution of Au atoms with Pd ones revealing a place exchange mechanism. Such interpretation is supported by experimental and theoretical observations of Cu/Pt(111) showing the appearance of substituted atoms in the topmost layer as dark spots in atomically resolved STM images [30].

The presence of similar substitution sites has been observed for extremely low coverage, while it was not possible to identify such process for coverage exceeding 0.012 ML. This is probably due to the high sticking coefficient at these sites, which causes the very rapid growth of an island covering the exchange site itself. It must be also considered that Pd does not show a very pronounced degree of intermixing with the gold atoms of the surface, at least at room temperature, since no sign of Au presence in Pd islands has been disclosed. Place exchange may be the result of a two stage process in which some of the Pd atoms are first adsorbed in hollow sites and finally incorporated in the topmost surface layer as substituted atoms .

The pinning of Pd atoms diffusing on the surface on top of substituted sites represents a trapping mechanism compatible with the growth dynamics observed experimentally. The binding energy of adatoms in those sites is expected to be much higher than for adsorbed atoms in gold hollow sites, so that a stable nucleus of aggregation is formed and the nucleation of islands proceeds quickly capturing Pd atoms diffusing nearby. If the sticking coefficient is high only in substituted sites, then clustered island growth may occur more easily with respect to the case of equally probable nucleation at each site. The replacement of gold with Pd is basically a stress-relief mechanism, so that the nucleation of islands is driven by the stress distribution on the surface. According to molecular dynamics simulations [17], the minimization of the surface energy leads to accumulation of a large stress at the elbows of the reconstruction, therefore these are the first sites to be occupied by islands. It is likely that a perfectly periodic stress distribution cannot be obtained on real reconstructed surfaces, so that nucleation first concentrates in regions, or rows of bulged/pinched elbows, exhibiting a slightly higher degree of strain. According to this picture, the complex rearrangements leading to irregularities of the reconstruction pattern seem to induce a lower degree of stress in the potential nucleation sites, resulting in a lower substitution probability.

### C. Surface – Overlayer Interaction

We complete our discussion focusing on the interaction between already formed palladium islands and the underlying Au(111) reconstructed surface (i.e. in the case of high coverage). The Pd island grown over a fcc region shown in fig. 3 exhibits a close-packed arrangement whose lattice constant exceeds the Pd bulk nearest-neighbour distance (2.75 Å) by 0.10 ± 0.04 Å, approaching very closely the atomic spacing of gold. This indicates that islands of several tens of atoms organize themselves in a pseudomorphic way with respect to the Au substrate, accumulating a rather large tensile stress until the atomic spacing approaches that of the substrate. This is quite interesting, since ridges display atomic spacing which is closest to that expected for Pd (2.75 Å). Indeed, according to first-principles simulations, the smallest atomic spacing in the $(22 \times \sqrt{3})$ cell is found to be 2.80 Å in correspondence of the discommensuration lines, while values of 2.87 Å and 2.85 Å are reported for fcc and hcp regions respectively [20]. In spite of that, islands systematically avoid growing on the ridges, and this tendency is maintained at increasing coverage, with substantial consequences on the surface morphology. This suggests that the minimum of energy for islands is represented by a close-packed isotropic arrangement, even though stressed, which can be obtained on fcc or hcp domains. It is also worth noticing that, according to a theoretical study of the structure of palladium clusters up to 10 atoms on Au(111) [31], no tensile stress is expected in such small aggregates, notwithstanding the 5% lattice mismatch. As a consequence, our results suggest that a critical size for the transition from non-commensurate to pseudomorphic growth may exist in between nearly 10 and a few tens of atoms.

Once the saturation regime is reached, island size is comparable to the distance between two adjacent ridges. Then the growth proceeds equally in all directions, without following the orientation of the ridges, as already pointed out in previous studies [1]. This reveals a competition between islands, which tend to maintain a uniform close-packed structure, and the underlying gold reconstruction. At 0.25 ML island growth causes distortion of the reconstruction, which is "pushed away" by islands entirely lying on fcc or hcp domains. Such behaviour is not common to all elements showing preferential nucleation on Au(111), for example Fe and Co islands grow without significantly influencing the periodic geometry of the surface [9], [32]. At this coverage the combined effect of island coalescence and the creation of new nucleation sites, due to the herringbone pattern distortion, brings to significant increase of the width of the island size distribution. At 0.50 ML the island-surface interaction brings to a considerable instability of the reconstruction, together with mass transport phenomena between neighbouring palladium islands, as revealed by the two STM images acquired in succession reported in Fig. 10. It is interesting to note that the occurrence of mass exchange is accompanied by the disappearing of a ridge separating the two islands. Mass transfer may be a mechanism for islands to release accumulated tensile stress, but its occurrence is energetically convenient between islands lying on the same type of close-packed domain. In fact the crossing of a ridge is expected to be energetically very unfavourable for islands, due to the change of lattice parameter and to the establishment of internal fcc/hcp mismatch. Nevertheless, at even higher coverage, the observation of discommensuration lines running underneath the islands is reported [1], suggesting that a very large accumulation of strain is established in the overlayer.

## IV. CONCLUSIONS

A scanning tunneling microscopy study of Pd island nucleation and growth on the Au(111) $22 \times \sqrt{3}$ reconstructed surface is here presented. At very low coverages, isolated Pd atoms are found at the elbows and distortions of the Au(111) reconstruction where they settle in hollow sites between gold atoms. At these preferential sites we observe substituted Pd atoms in lattice site position with respect to gold atoms of the surface, with an initial prevalence of the fcc regions outside the bulged elbows. Our experimental observations suggest that substituted atoms act as nucleation centres for island growth, thanks to the high sticking coefficient between Pd atoms. Increasing the coverage, preferential sites of island nucleation are filled following a precise course (first bulged elbows, then pinched elbows and at last distortions) until the island density reaches the density of available nucleation sites (saturation regime). This behaviour may be explained by the fact that substitution is promoted only in correspondence of localized sites of the Au(111) topmost layer, where particular stress conditions are verified. Therefore we propose substitution (i.e. place exchange) in preferential sites as the starting mechanism for ordered nucleation of Pd islands on Au(111) surface. Pd atom adsorption in hollow sites of the Au(111) surface plays also a role in favouring atom aggregation nearby substitution sites.

Islands are shown to grow in a pseudomorphic way on the Au(111) surface, displaying a higher lattice parameter with respect to that of bulk Pd. The tensile stress can be transferred to the underneath surface leading to reorganization of the underlying reconstruction and mass transport phenomena between Pd islands.

**FIGURE CAPTIONS**

Fig. 1.

STM images of the clean Au(111) surface. (a) Atomically resolved STM image (V=-0.4V, I=0.6nA) showing the (22 × $\sqrt{3}$ ) unit cell (indicated by a dashed rectangle). The bright ridges (corrugation ~0.02 nm) running along the $\langle 11\bar{2} \rangle$ direction represent the discommensuration lines composed by atoms in between two different staking positions. (b) Large scale STM image (-1V, 1nA) of the typical "herringbone" surface reconstruction of Au(111).

Fig. 2.

STM images of Pd deposition on Au(111) at 300 K with increasing coverage. (a) 0.012 ML, *V*=-2V, I=0.1 nA; (b) 0.026 ML, *V*=-1.4V, I=0.1 nA; (c) 0.035 ML, *V*=-0.6V, I=2 nA; (d) 0.14 ML, *V*=-2V, I=1 nA.

Fig. 3.

Pd island at atomic resolution grown on a fcc domain near a bulged elbow. Pd coverage is 0.026 ML (as in fig.2-b), tunneling conditions are *V*=1V, I=0.2 nA. Line profile shows a one-layer high Pd island.

Fig. 4.

Island size distributions for coverage ranging from 0.012 to 0.14 ML, corresponding to the images reported in fig.2. Points are fitted with gaussian functions.

Fig. 5.

Log-log plot of the island density *n* (top) and of the mean island size <*s*> (bottom) as a function of the coverage *θ*. Cross over form pre-saturation to saturation regimes is indicated by a dashed line.

Fig. 6.

STM images (*V*=1.5V, I=0.9nA) of the same 20x20 nm$^2$ area, acquired in succession after deposition of 0.026 ML. Arrows indicate some missing or shifted features. The time interval between the two images is 75 sec.

Fig. 7.

STM image ($V$=1.5V, I=0.9nA) of a Pd atom adsorbed in a hollow site of the Au(111) surface. Black dotted lines are a guide for the eye and follow the directions of the regular rows of Au(111) atoms. Line profile crossing the adsorption site (white arrow) is also reported showing a depression of 0.3-0.4 Å.

Fig. 8.

STM image ($V$=-2V, I=0.1nA) at low Pd coverage (0.012 ML). Adsorbed Pd atoms are distributed nearby all nucleation sites: regular (both pinched and bulged elbows) and distortions without forming islands.

Fig. 9.

STM image ($V$=-0.2V, I=0.5nA) at extremely low Pd coverage (0.003 ML). Substitution of a few Pd atoms in lattice sites of the Au(111) surface is shown.

Fig. 10.

STM images of the same 70x55 nm$^2$ area, acquired in succession after a deposition of 0.5 ML of Pd. Black circles show some modifications in the Au(111) reconstruction, while black arrows indicate the direction of island-mass transport phenomena that have been take place between the two measurements. The time interval between the two images is 75 sec.

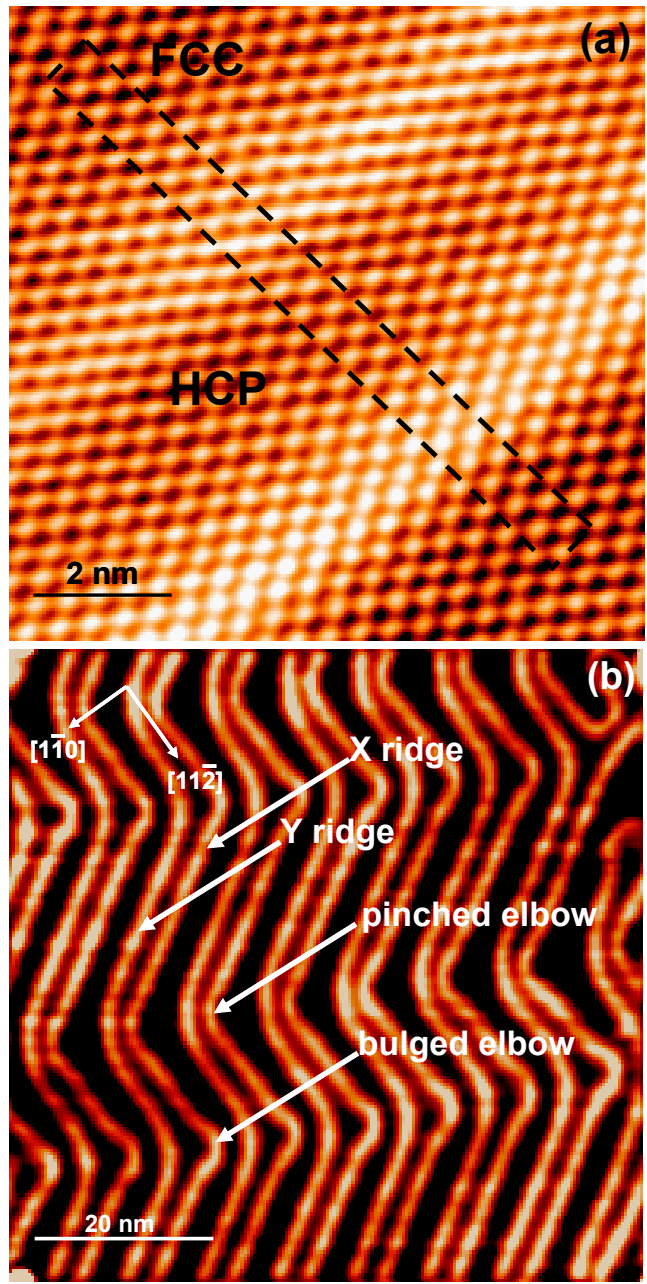

Figure 1

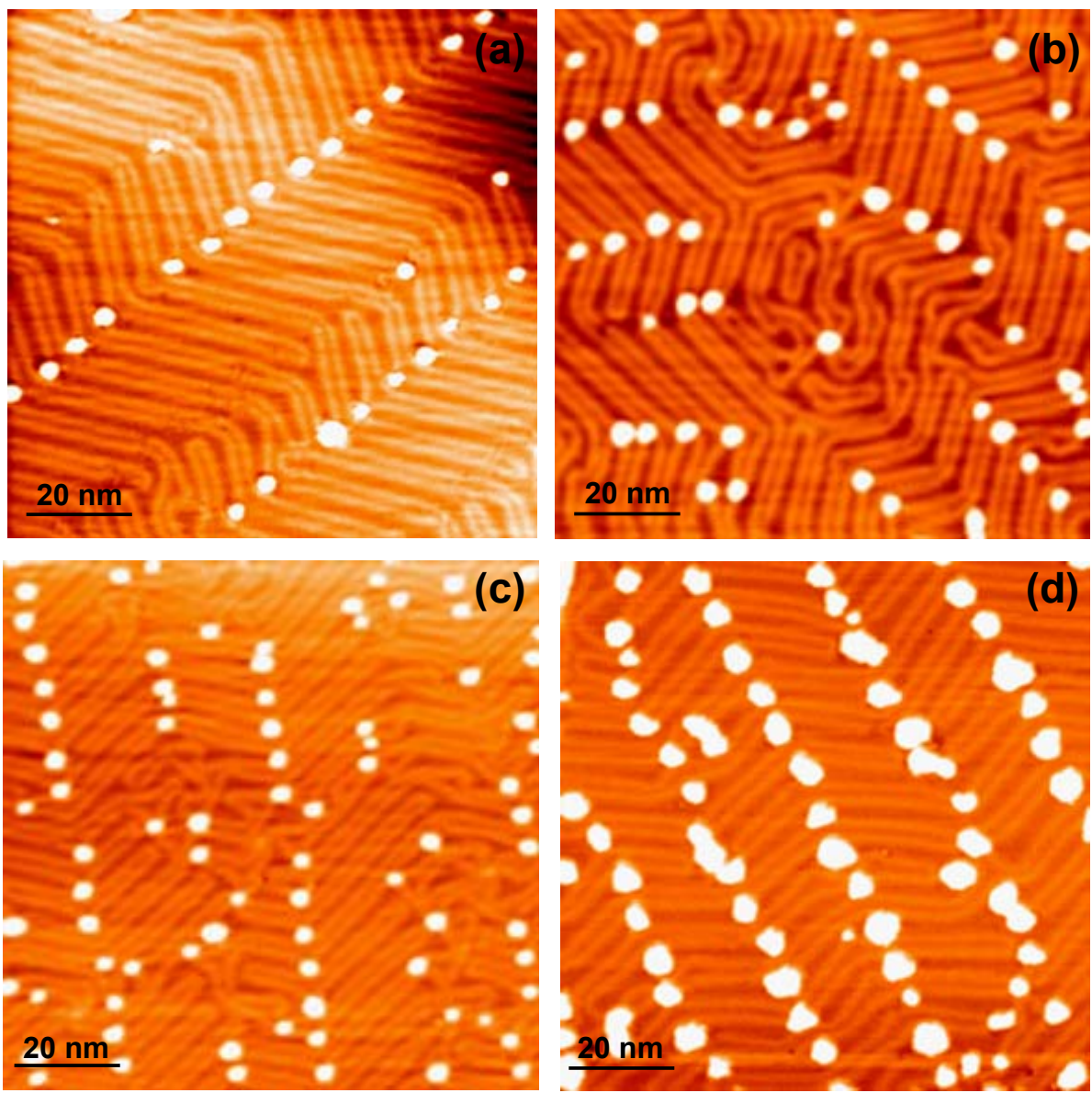

Figure 2

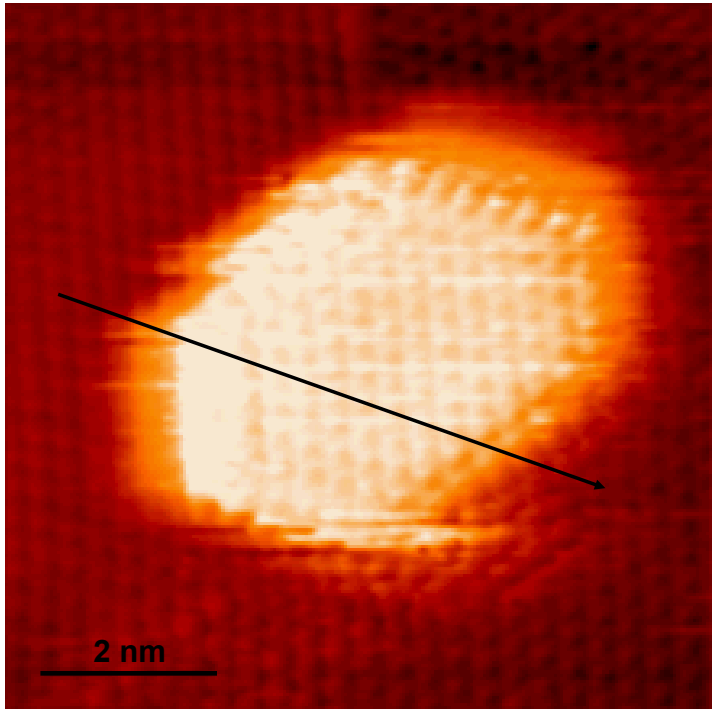
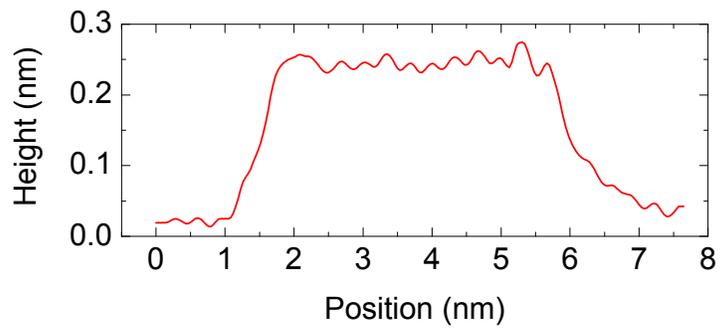

Figure 3

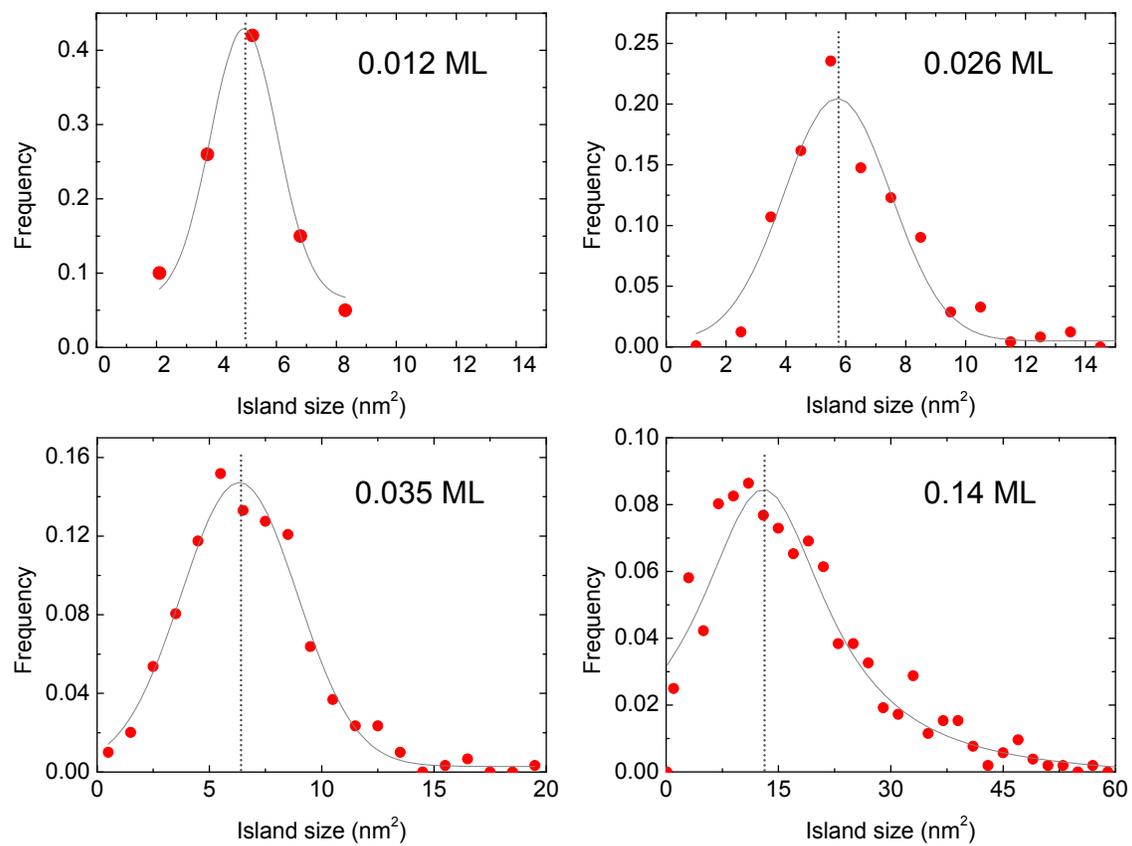

Figure 4

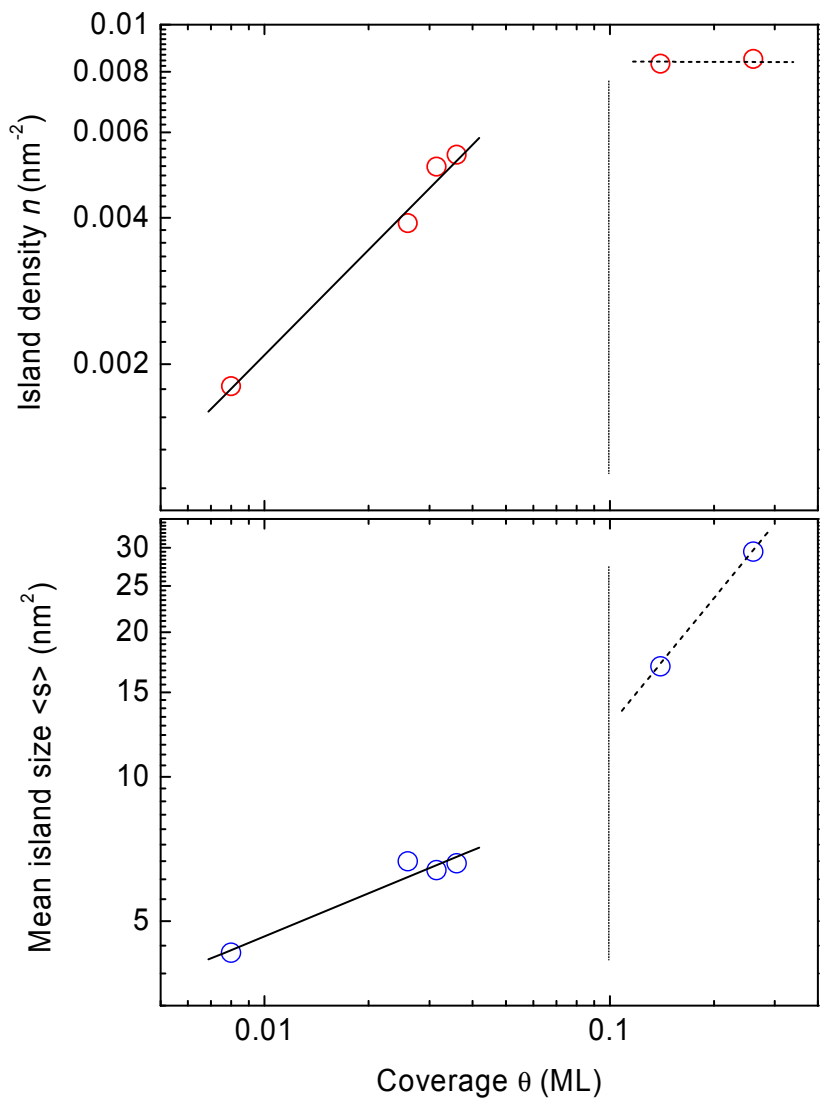

Figure 5

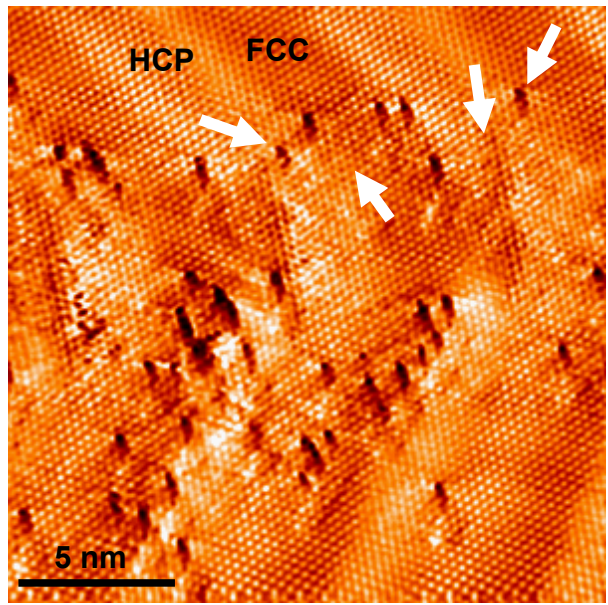
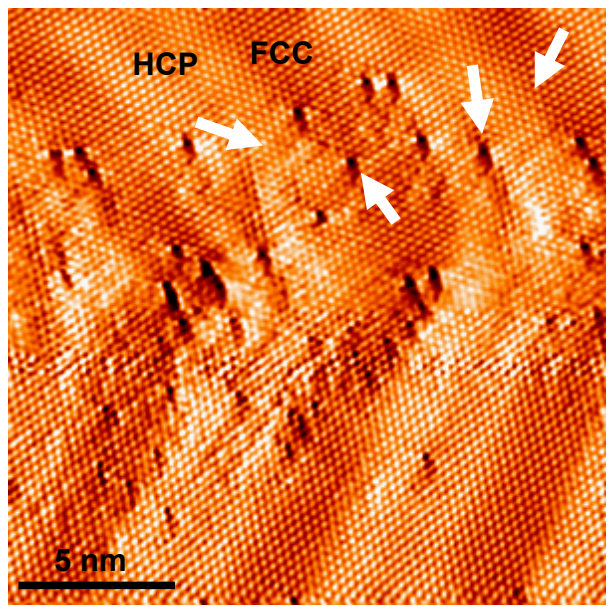

Figure 6

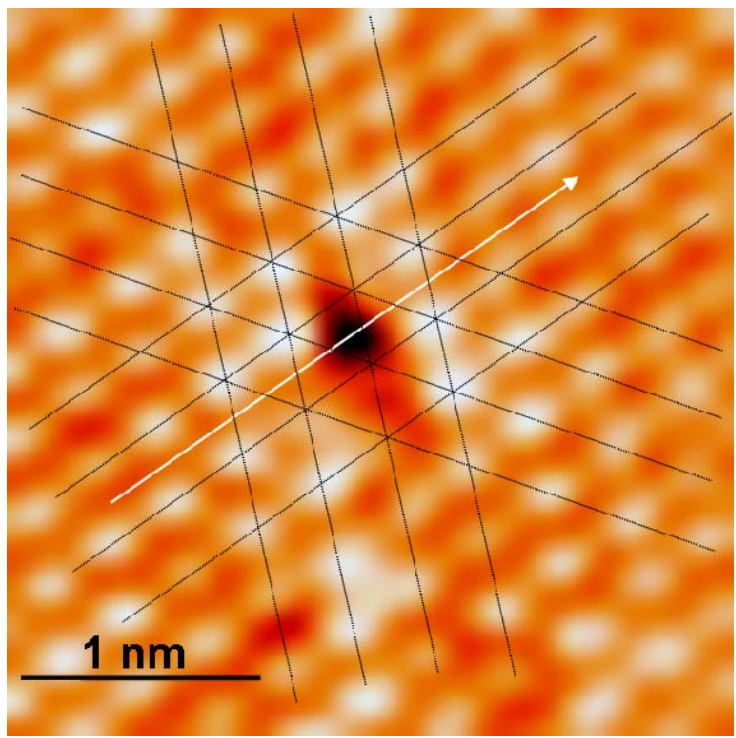
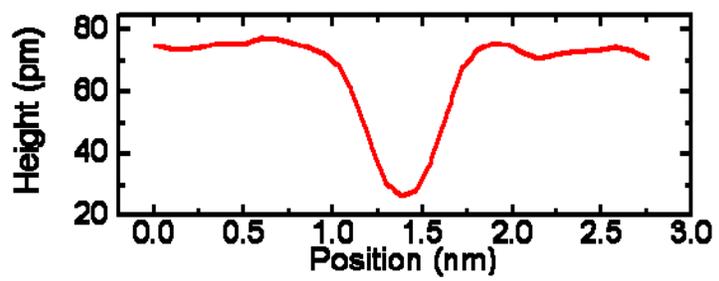

Figure 7

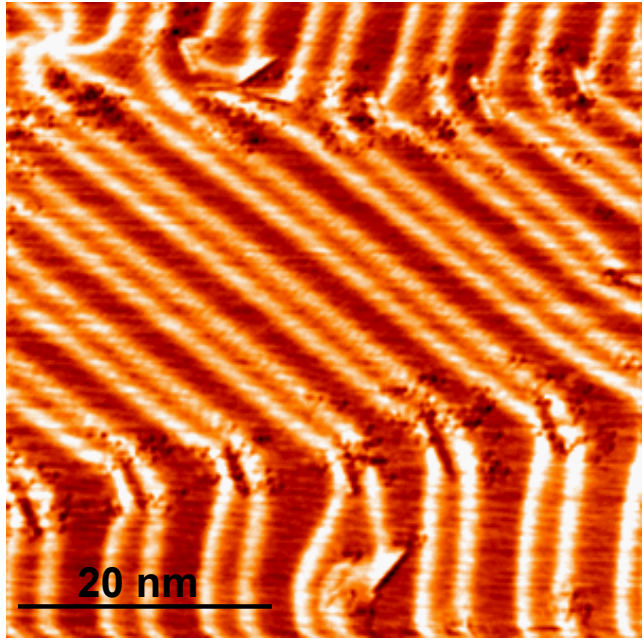

Figure 8

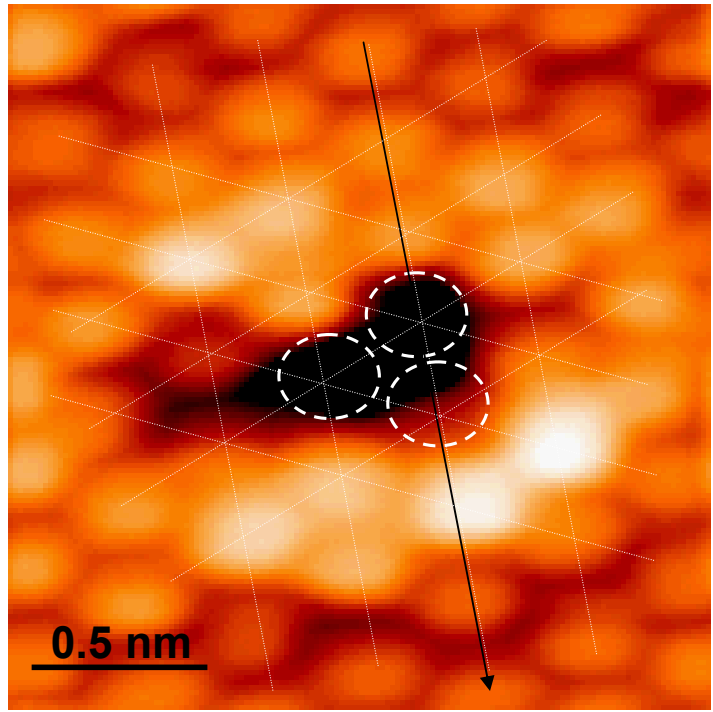
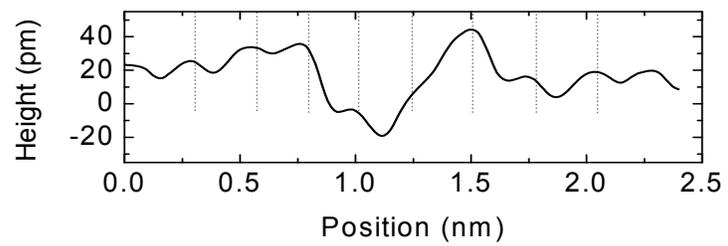

Figure 9

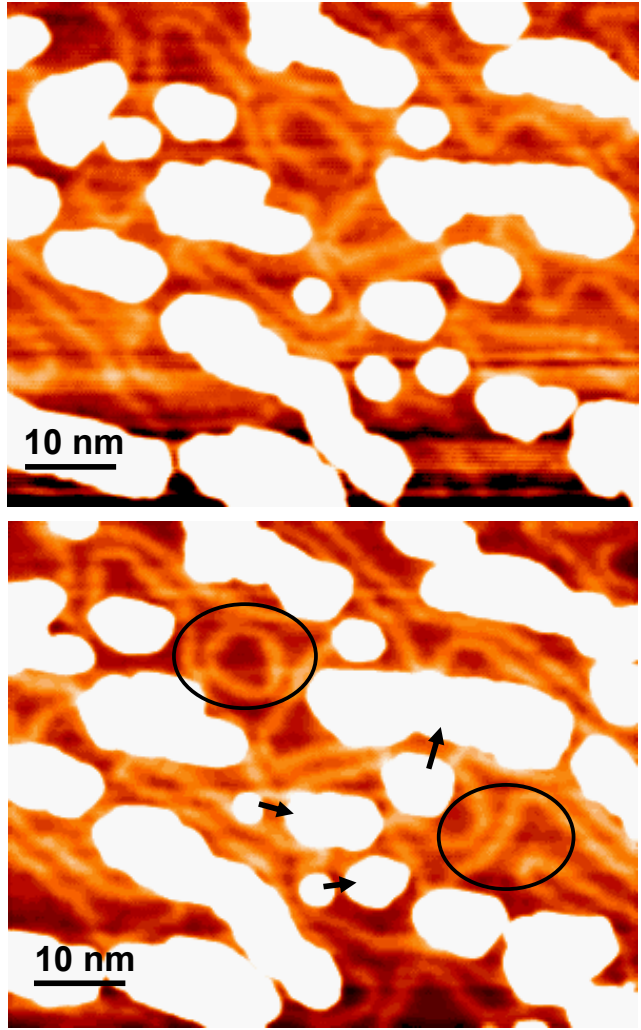

Figure 10